\documentclass[epj-spec-noh]{svjour}
\usepackage{graphics}

\begin{document}
\title{An ARPES view on the high-$T_c$ problem: \\phonons \textit{vs} spin-fluctuations}
\author{A. A. Kordyuk\inst{1,2}\fnmsep\thanks{\email{kordyuk@gmail.com}} \and V. B. Zabolotnyy\inst{1} \and D. V. Evtushinsky\inst{1}
\and D. S. Inosov\inst{3} \and T.~K.~Kim\inst{1} \and B.~B\"{u}chner\inst{1} \and S.~V.~Borisenko\inst{1}}
\institute{IFW Dresden, P.O. Box 270116, D-01171 Dresden, Germany \and Institute of Metal Physics of National Academy of Sciences of Ukraine, 03142 Kyiv, Ukraine \and Max Planck Institute for Solid State Research, D-70569 Stuttgart, Germany}
\abstract{
We review the search for a mediator of high-$T_c$ superconductivity focusing on ARPES experiment. In case of HTSC cuprates, we summarize and discuss a consistent view of electronic interactions that provides natural explanation of both the origin of the pseudogap state and the mechanism for high temperature superconductivity. Within this scenario, the spin-fluctuations play a decisive role in formation of the fermionic excitation spectrum in the normal state and are sufficient to explain the high transition temperatures to the superconducting state while the pseudogap phenomenon is a consequence of a Peierls-type intrinsic instability of electronic system to formation of an incommensurate density wave. On the other hand, a similar analysis being applied to the iron pnictides reveals especially strong electron-phonon coupling that suggests important role of phonons for high-$T_c$ superconductivity in pnictides.
} 
\maketitle
\section{Introduction}
\label{intro}

The problem of high temperature superconductivity \cite{Ginzburg} has been opened for experimental study by the discovery of high-$T_c$ cuprates back in 1986 but among a number of models suggested in the subsequent years none has been generally accepted so far. In this situation it is reasonable to ask what are the chances for resolving the problem in the near future and what are the reasons for reading yet another review on this subject. We believe, it is the present level of experimental techniques, developed, to the large extent, in the endeavor to solve the high-$T_c$ problem, that suggests positive answers to the both questions. At the present time, it seems that both the experimental accuracy and understanding of how to decipher the key interactions from the experimental data have just reached the complexity level of the problem.

In particular, the angle resolved photoemission spectroscopy (ARPES), continuously improving \cite{ARPES}, nowadays provides a direct view of the rich spectrum of one-particle fermionic excitations, which encapsulates all the interactions of the electrons in crystal, with the accuracy better than 0.25 \% of the Brillouin zone (0.004 {\AA}$^{-1}$) in momentum \cite{KoralekRSI2007}, a few meV in energy \cite{KissRSI2008,IshizakaPRB2008}, and down to 1 K sample temperature \cite{BorisenkoLFA,KordyukLFA}. On the other hand, the progress in understanding the structure of ARPES spectra is proven by the successful bridging ARPES to other experimental techniques such as inelastic neutron scattering (INS) \cite{ChatterjeePRB2007,InosovPRB2007,DahmNP2009}, scanning tunneling spectroscopy (STS) \cite{McElroyPRL2006,ChatterjeePRL2006,KordyukJES2007}, Raman spectroscopy \cite{Raman}, as well as to the macroscopic probes, such as resistivity and Hall measurements \cite{EvtushinskyTaSe2}, $\mu$SR \cite{EvtushinskyNJP}, etc. So, the evaluation of the existing models for consistency with different experimental probes as well as their reevaluation with the improved experimental accuracy appears timely and should be important for establishing the true mechanism of high-$T_c$ superconductivity.

In this paper, we summarize a consistent view of electronic interactions in HTSC cuprates, in which, the spin-fluctuations play a decisive role in formation of the fermionic excitation spectrum in the normal state and are sufficient to explain the high transition temperatures to the superconducting state while the pseudogap phenomenon is a consequence of a Peierls-type intrinsic instability of electronic system to formation of the spin density waves. We evaluate the robustness of our conclusions and room for the electron-phonon interaction in future experiment. Ironically, a similar analysis being applied to the iron pnictides reveals especially strong electron-phonon coupling that suggests important role of phonons in high-$T_c$ superconductivity of pnictides.

\section{Role of spin-fluctuations in cuprates}
\label{sec:2}

At this point, it is reasonable to ask whether searching for a mediator or a `pairing glue' \cite{AndersonSci2007} is relevant for the superconductivity in cuprates at all? Leaving a detailed answer to this question for a future study we note that spectroscopically, i.e. from the point of view of one-particle excitation spectrum, known with present experimental accuracy, the superconducting state of the cuprates reveals nothing\footnote{We do not consider the issue of Fermi-liquidity here but the pseudo-gap phenomenon is discussed in Sec.~\ref{sec:3}.} beyond the conventional BCS model where both the spin-fluctuations and phonons have provided numerous evidences for the role of the `pairing glue'. Therefore, here we focus on spectroscopic differences that can help to choose unambiguously between those two scenarios and stay open for any inconsistencies of the one-particle spectral function with BCS theory, hoping that, if the mechanism of superconductivity is principally different, the difference can be eventually identified by ARPES.

Starting from the first proposals and up to now the main argument between the promoters of either phonons or spin-fluctuations against the opposite scenario is that the coupling strength of the relevant excitations to electrons is by far not sufficient to provide the high-$T_c$ pairing \cite{KeePRL2002,AbanovPRL2002,Giustino}. We do not discuss this kind of arguments here but approach the problem from the empirical side, trying to formulate a set of critical experimental observations which can be described by one scenario but not by the other.

The idea of this approach is to understand the constituents of the quasiparticle spectrum of cuprates in the normal state, identify all the essential interactions which form this spectrum and estimate the strength of their coupling to electrons. Naturally, one should aim at a complete understanding of the quasiparticle spectrum in the whole Brillouin zone (BZ) but even its simple mapping with a sufficient accuracy requires tremendous experimental efforts. Therefore, as the first challenge, a successful model should be able to describe the main peculiarities of this spectrum known as `nodal kink' and `antinodal dip', reasonably explaining their behavior with doping and temperature. And since the key requirement for the empirical approach is experimental accuracy, we focus on two compounds: bilayer Bi(Pb)$_2$Sr$_2$CaCu$_2$O$_{8+\delta}$ (BSCCO), the most `arpesable' high-$T_c$ compound \cite{KordyukLTP2006}, and YBa$_2$Cu$_3$O$_{7-\delta}$ (YBCO), the best compound to compare ARPES and neutron scattering experiments \cite{DahmNP2009}.

\subsection{Antinode}
\label{sec:2.1}

\begin{figure}
\centering
\resizebox{0.52\columnwidth}{!}{%
  \includegraphics{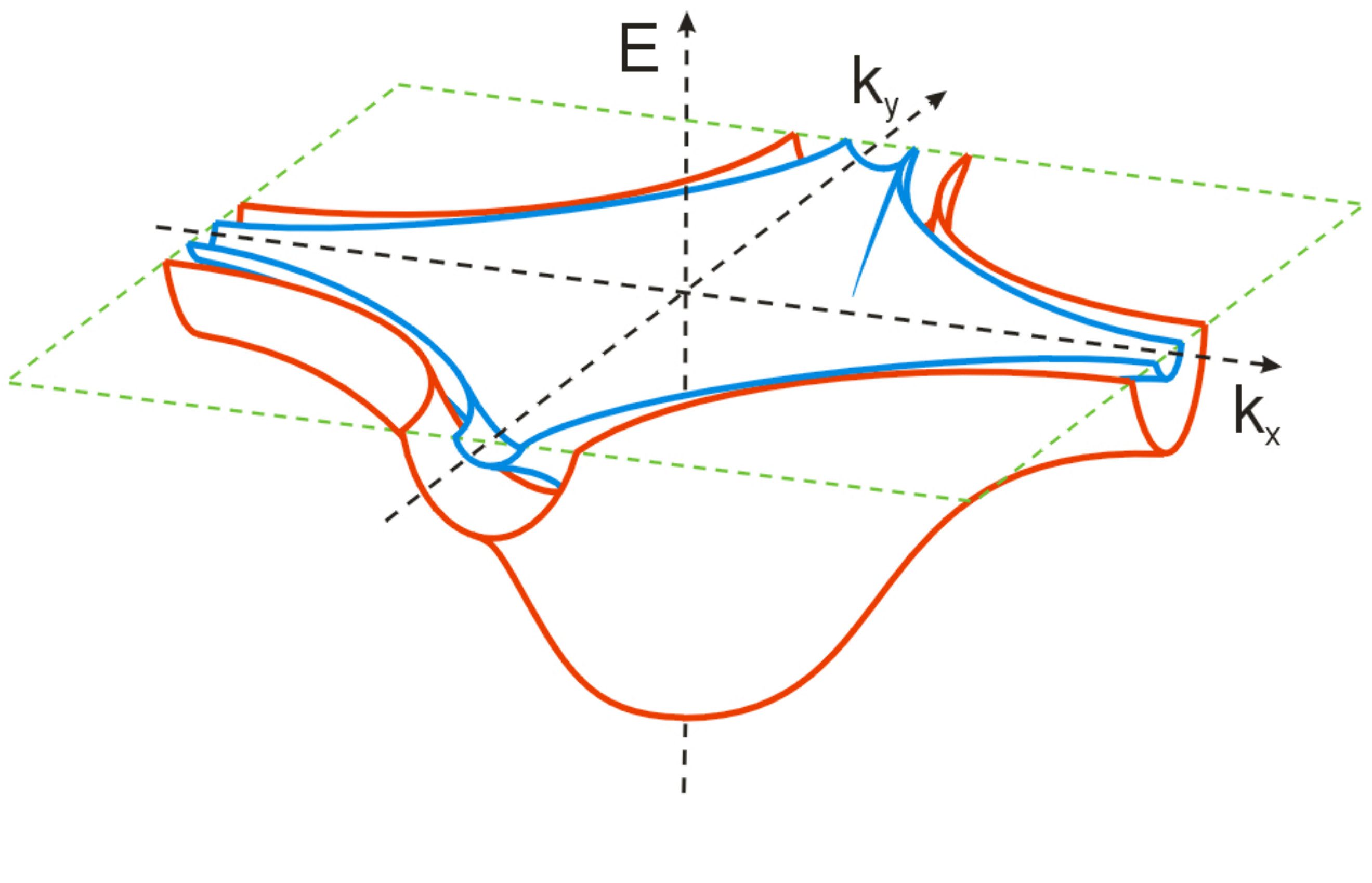}}
\resizebox{0.37\columnwidth}{!}{%
  \includegraphics{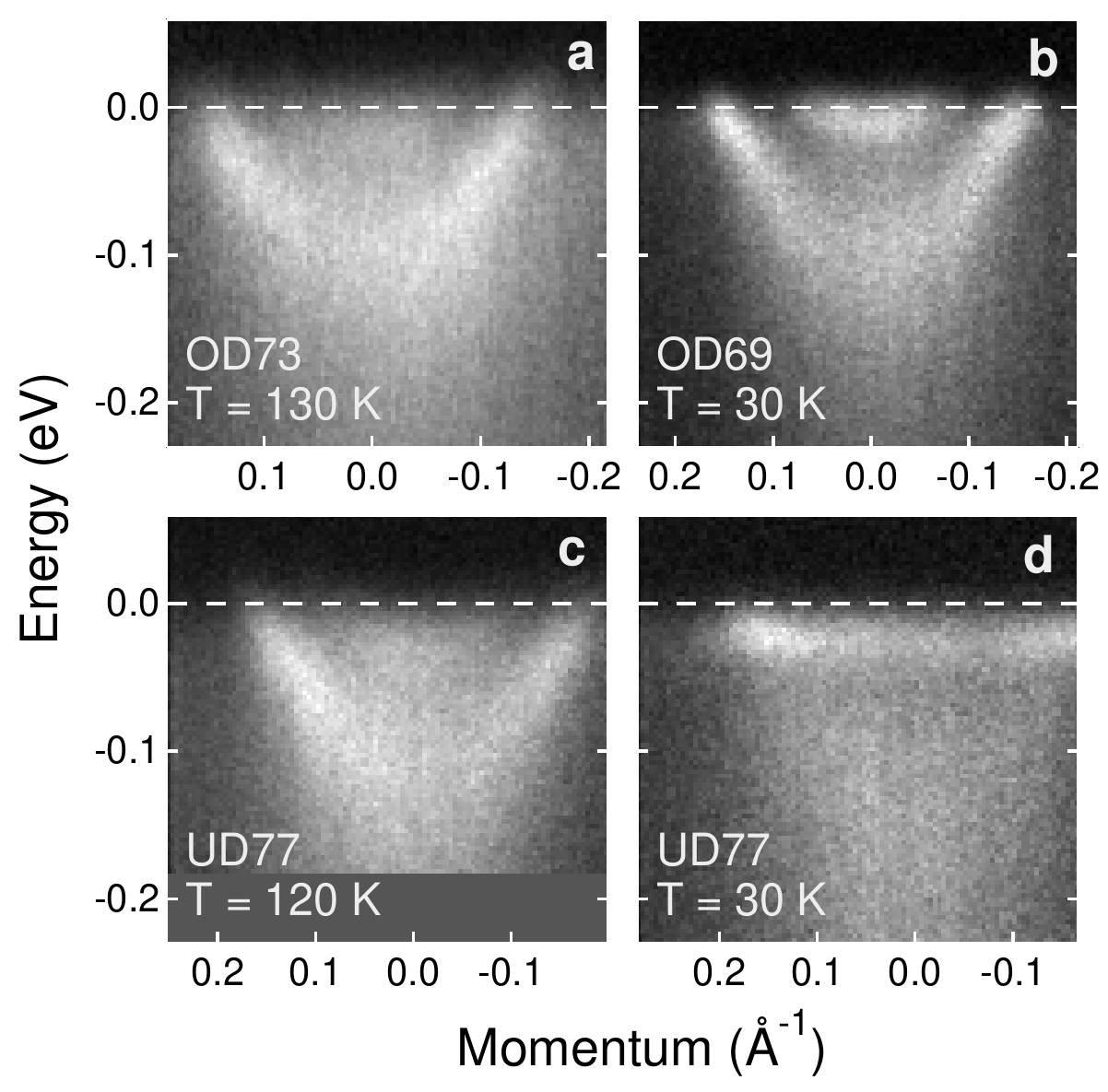} }
\caption{(left) Occupied low energy electronic band structure of the bi-layer high-$T_c$ cuprates is sketched in the first Brillouin zone. (a-d) ARPES spectra from the antinodal cut trough the saddle point illustrate strong dependence of the renormalization strength on doping and temperature. After \cite{BorisenkoPRL2003}.}
\label{fig:1}       
\end{figure}

The expected conduction band structure of the bi-layer cuprates in the first BZ is sketched in Fig.~1. The areas of the primary interest, the nodal  and antinodal regions, lie along the BZ diagonal and around the Fermi surface crossings by the BZ boarder, respectively. The most outstanding feature of the quasiparticle spectrum has been observed on the energy distribution curves (EDC) from the antinodal region back in 1991 \cite{DessauPRL1991} and known as the `peak-dip-hump' lineshape. Considering the `dip' as a consequence of very strong scattering of the electrons by a `bosonic mode' immediately nominates the appropriate bosons for the role of the pairing glue. So, the understanding of its mechanism is called for.

Taking into account the role of the bilayer splitting, which has appeared to be the main responsible for the peak-dip-hump lineshape in the overdoped BSCCO \cite{KordyukPRL2002}, the true doping and temperature dependence of the coupling to the mode has been revealed: it emerges below $T_c$ but its strength decreases monotonically with overdoping \cite{BorisenkoPRL2003,GromkoPRB2003}, vanishing at about 24\% of holes per unit cell \cite{KimPRL2003}. The examples of the ARPES spectra taken along the antinodal direction in Fig. 1 (a)-(d) illustrate this. The energy scale and location in momentum of this strong renormalization, as well as its abrupt emergence below $T_c$, point unambiguously to the famous magnetic resonance \cite{EschrigPRB2003,ChubukovPRB2004,EschrigAiP2006} as the main scatterer. However, the observed strong doping dependence of the renormalization strength seems to be the most crucial evidence for the spin-fluctuation scenario against phonons. Within the former scenario, the strong doping dependence can be naturally understood as a proximity to the antiferromagnet (see Fig. 2(d)), while the complete vanishing of renormalization with overdoping is hard to reconcile with phonons.

\subsection{Node}
\label{sec:2.2}

Exactly the opposite argument has been applied to the nodal direction where an ubiquity of the kink \cite{VallaSci99} in the quasiparticle dispersion for different compounds, doping levels, and temperatures has been considered as an evidence for the strong coupling to a phonon mode \cite{LanzaraNature01}. One should note, however, that since the renormalization strength along the nodal direction is much smaller than in the antinodal region, its parametrization requires a much more careful analysis. In other words, in order to conclude about the properties of the bosonic contribution to quasiparticle spectrum one should single it out first. Practically this means that the electron-boson part, $\Sigma_{b}$, should be singled out from the total quasiparticle self-energy, which, in turn, should be derived from the ARPES spectrum and disentangled from the artificial effects such as bilayer splitting \cite{KordyukPRB2004}, superstructure, photoemission matrix elements, etc. As a result, while the nodal experimental dispersion looks similar for different compounds, doping levels and temperatures (see \cite{LanzaraNature01,ZhouNature03} or Fig. 2(a)), the derived $\Sigma_{b}$ varies essentially and, when the experimental artifacts are properly accounted for, exhibits critical dependence on doping and temperature \cite{KordyukPRL2004,KordyukPRB2005,KordyukPRL2006}.

\begin{figure}
\center{
\resizebox{0.56\columnwidth}{!}{%
  \includegraphics{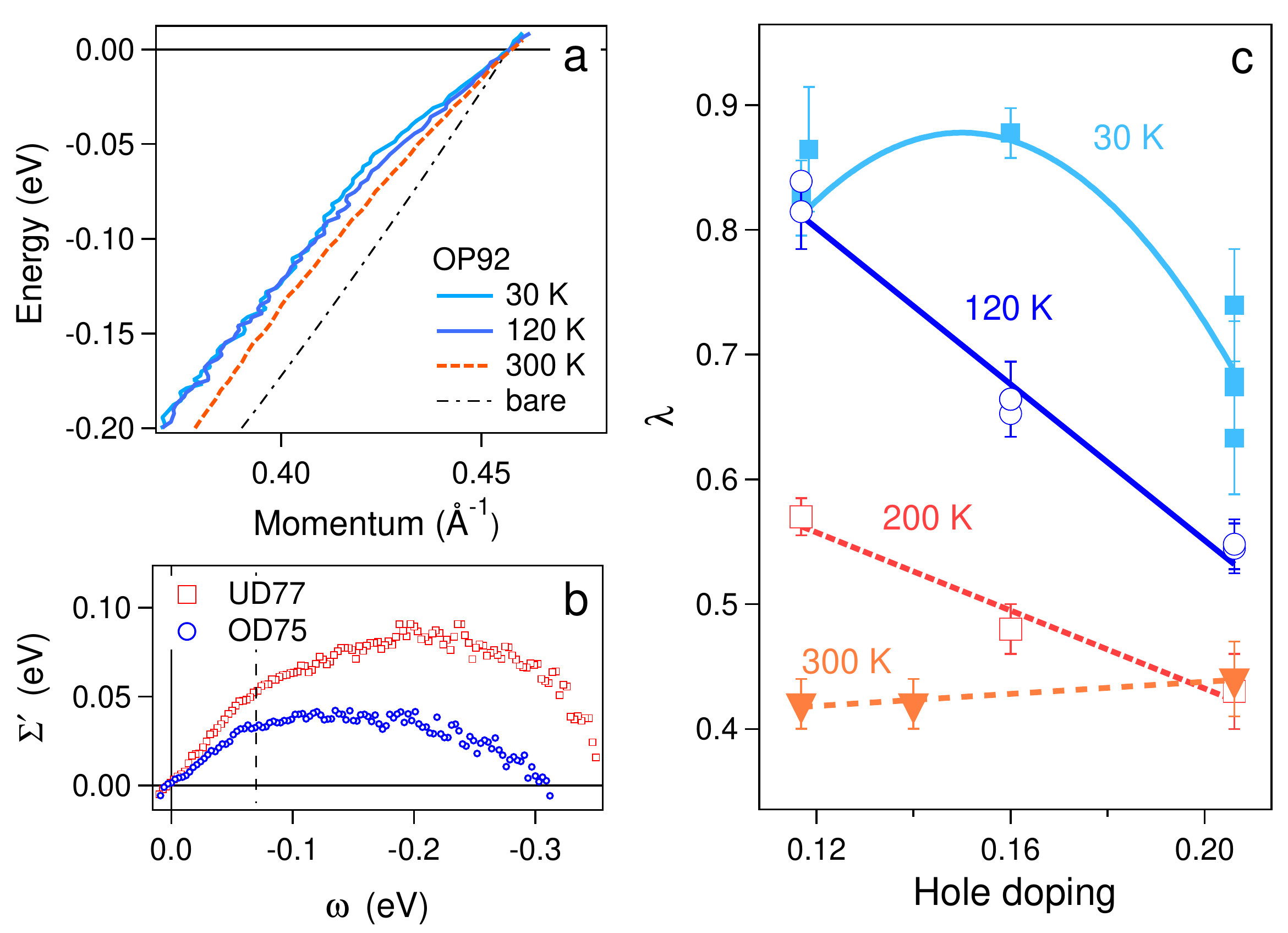} }
\resizebox{0.37\columnwidth}{!}{%
  \includegraphics{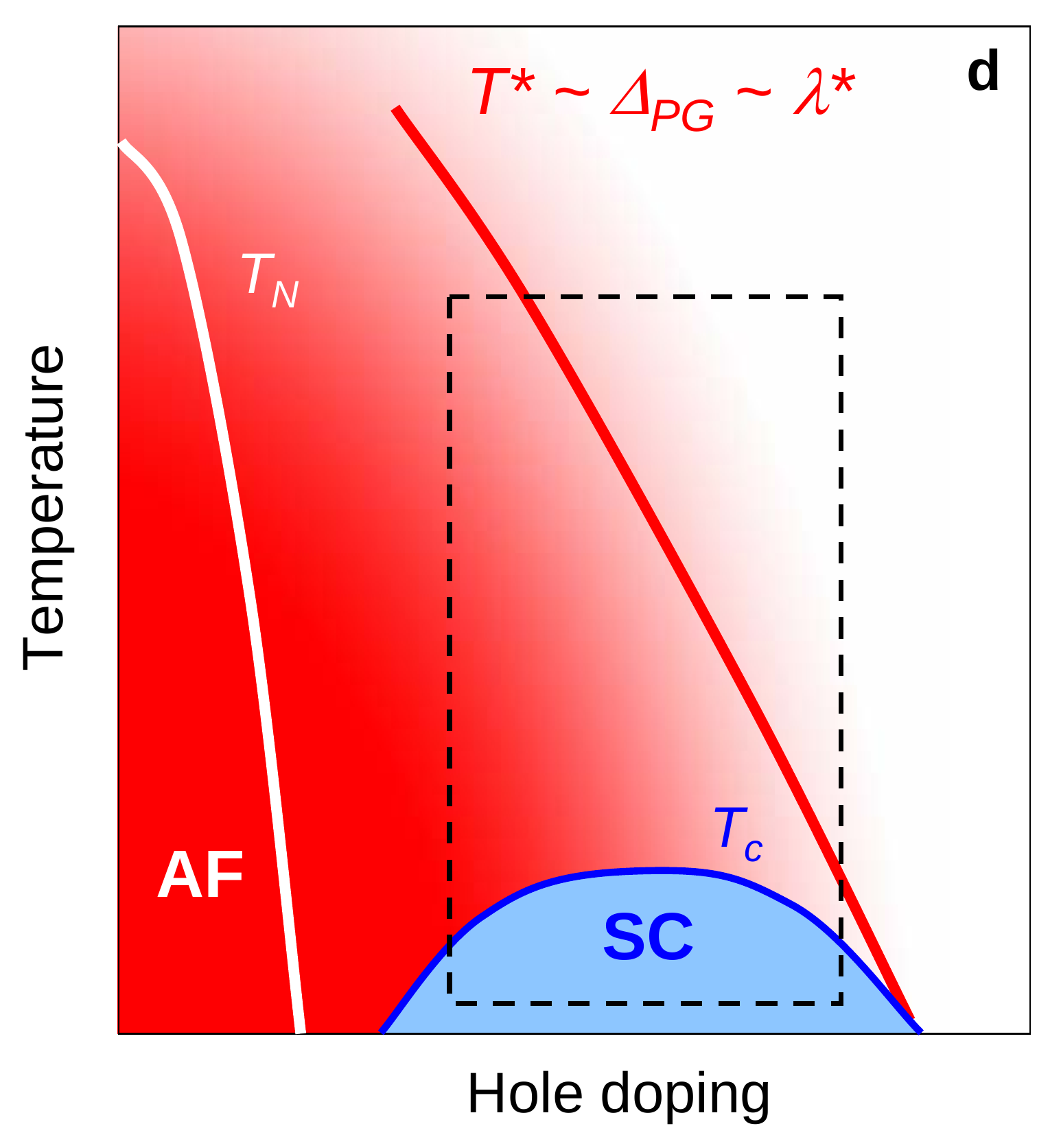} }}
\caption{Temperature and doping dependence of renormalization along the nodal direction. (a) MDC dispersions of an optimally doped BSCCO ($T_c = 92$K) at different temperatures \cite{KordyukPRL2006}. (b) Real part of the self-energy for the overdoped (OD 75K) and underdoped (UD 77K) samples \cite{KordyukPRB2005}. (c) Dependence of the total coupling strength on doping and temperature \cite{KordyukPRL2006}. (d) A sketch to illustrate similarly strong $xT$-dependence of $T^*$, pseudogap, and coupling strength in terms of proximity to antiferromagnet: $\lambda^*$ refers to both the nodal electron-phonon coupling strength, $\lambda_b$, in the pseudogap state and antinodal renormalization strength below $T_c$; $\lambda_b(x, T<T_c)$ is also illustrated by the color gradient.}
\label{fig:2}       
\end{figure}

In Ref. \cite{KordyukPRL2006} both the nodal quasiparticle self-energy and the bare band dispersion have been derived from ARPES spectra with a self-consistent procedure \cite{KordyukPRB2005}. An example of temperature dependence of the experimental dispersion is shown in Fig. 2(a) for the optimally doped BSCCO while Fig. 2(b) illustrates the doping dependence of the real part of the self-energy. The data suggests that the total self-energy can be considered a sum of three components: $\Sigma(\omega, T, x) = \Sigma_{imp}(T, x) + \Sigma_{el}(\omega) + \Sigma_{b}(\omega, T, x)$. Here $\Sigma_{imp}$ stands for elastic scattering on impurities that generally increases with $T$ and is sample- rather than doping-dependent \cite{EvtushinskyPRB2006}. $\Sigma_{el}$ has been associated with the electron-electron scattering due to Coulumb interaction \cite{KordyukPRL2004}. It is important to stress that the electron-electron scattering in HTSC cuprates makes an essential contribution to the low energy bare band renormalization with $\lambda_{el} \approx 0.5$.\footnote{Here we define the coupling strength as $\lambda = -\Sigma'(\omega)/\omega$ at $\omega \rightarrow 0$.} This component almost does not depend on $T$ and $x$ but is a smooth function of $\omega$, though, in terms of its real part, with the maximum on the scale of the band width \cite{KordyukPRB2005}. The last contribution, $\Sigma_{b}$, is left for the electron-boson coupling. Its imaginary part is a step-like function, and it is this component which is only responsible for the kink in the dispersion at 50--80 meV binding energy (depending on cuprate family, doping and temperature). What is important, $\Sigma_{b}$ depends critically on $T$ and $x$, following the same trend on the phase diagram associated with the `proximity to antiferromagnet' (see the sketch in Fig. 2(d)). Fig. 2(c) shows the total coupling strength, $\lambda$, as function of $x$ and $T$ that can be summarized as follows: $\lambda(x,T) = \lambda_{el} + \lambda_b(x,T)$. So, both the coupling strength to the bosonic mode in the antinodal region below $T_c$ and the electron-boson coupling strength along the nodal direction just above $T_c$ shows the same trend with doping, as indicated by the solid red line in Fig. 2(d). We consider such a dependence as the most robust evidence in favor of the spin-fluctuation rather than phonon origin of the nodal kink.

The revealed $xT$-dependence of the nodal and antinodal renormalizations leaves some open questions. The most straightforward one is the `kink puzzle'. Namely, why the antinodal renormalization disappears abruptly above $T_c$, while the nodal kink survives up to much higher temperatures? It has been suggested that the nodal kink is the result of coupling to rather persistent high energy part of the spin-fluctuation spectrum known as a gapped continuum \cite{ChubukovPRB2004,KordyukPRL2006}. To clarify this issue, the momentum dependence of both the quasiparticle and bosonic spectra should be taken into account.

\subsection{Bridging ARPES and INS}
\label{sec:2.3}

\begin{figure}
\center{
\resizebox{0.6\columnwidth}{!}{%
  \includegraphics{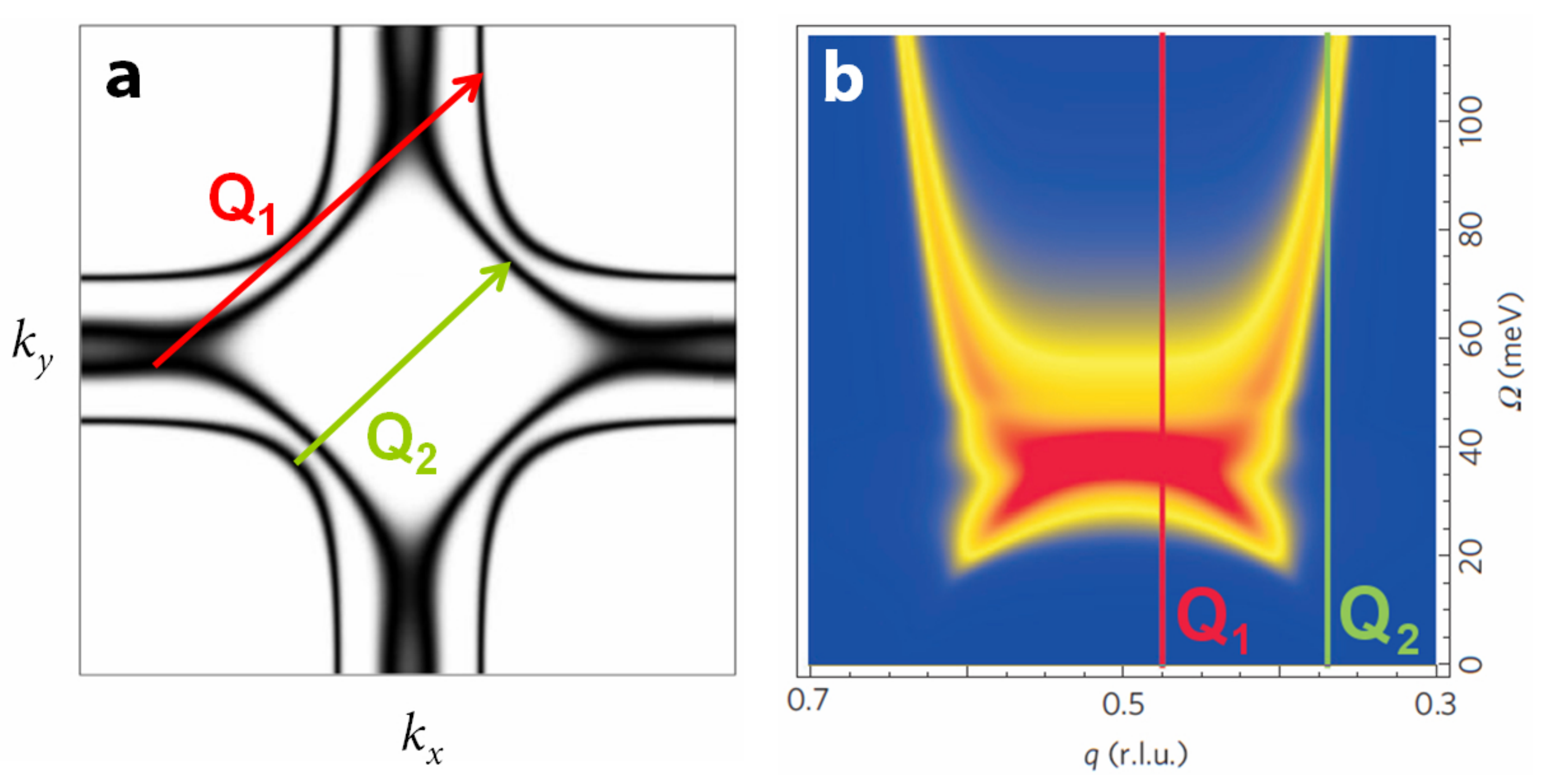} }
\resizebox{0.31\columnwidth}{!}{%
  \includegraphics{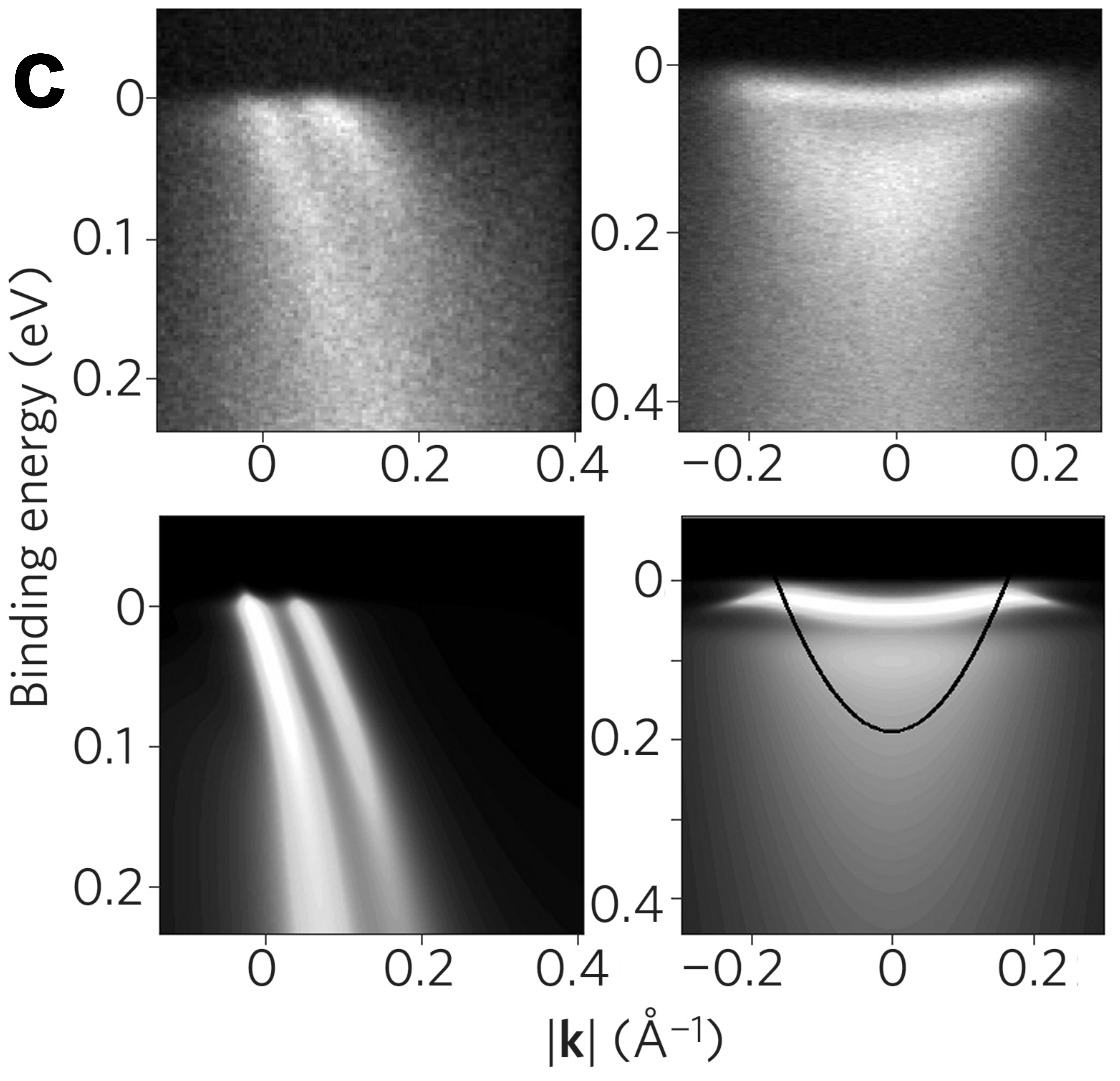} }}
\caption{Bridging ARPES and INS. (a) The Fermi surface of YBCO in the 1st BZ derived from ARPES data \cite{ZabolotnyyPRB2007} represents the fermionic Green's function. (b) The intensity of spin excitations along $Q = q(2\pi,2\pi)$ resulting from numerical fits to the INS spectra measured by V. Hinkov and B. Keimer (MPI, Stuttgart) \cite{DahmNP2009}. (c) Comparison of experimental (upper row) and theoretical (lower row) fermionic spectra (see Ref.  \cite{DahmNP2009} for details), by T. Dahm (University of T\"{u}bingen).}
\label{fig:3}       
\end{figure}

The aim of this section is to show that the spin-fluctuation spectrum, $\chi(\Omega, \textbf{q})$, is indeed consistent with the spectrum of one-particle fermionic excitations, $A(\omega, \textbf{k})$ $\propto$ Im($G$), in the whole reciprocal space. We start from the final empirical conclusion about the structure of the fermionic Green's function of the cuprates that can be formulated by a conceptual equation

\begin{equation}\label{E1}
G^{-1} = G^{-1}_0 - \bar{U}^2 G \star G \star G,
\end{equation}
which is the Dyson equation $G^{-1} = G^{-1}_0 - \Sigma$ extended by the following definition of fermionic self-energy and electronic susceptibility
\begin{eqnarray}\label{E2}
\Sigma &=& \bar{U}^2 \chi \star G,\\
\chi &=& G \star G.
\end{eqnarray}
The sense of these formulas is that the Green's functions of bare electrons and fermionic quasiparticles, $G_0$ and $G$, are related by a single parameter, $\bar{U}$, a spin-fermion coupling constant. The `$\star$' signs denote here the operations with the meaning of cross-correlation, but the exact relations for $\Sigma$ and $\chi$ can be found in Refs. \cite{DahmNP2009} and \cite{InosovPRB2007}, respectively. The justification of Eq. \ref{E1} can be separated, consequently, into two steps, according to Eqs.~2 and 3.

The first step is a search for `fingerprints' of the bosonic spectrum in the fermionic one. It can be also formulated as $G^{-1} = G^{-1}_0 - \bar{U}^2 \chi \star G$. Deriving $G(\omega, \textbf{k})$ and $G_0(\omega, \textbf{k})$ from ARPES and $\chi(\Omega, \textbf{q})$ from INS, one can check their mutual consistency and, if it is the case, estimate $\bar{U}$. Such calculations have been performed by T. Dahm \cite{DahmNP2009} on the basis of ARPES and INS spectra (see Fig. 3) measured for YBCO crystals from the same batch. The details of ARPES on YBCO can be found elsewhere \cite{ZabolotnyyPRB2007}. The overall similarity of experimental and calculated fermionic spectral functions, shown in Fig. 3(c), demonstrates clearly that the spin fluctuations can explain all the peculiarities of the electronic scattering in cuprates including its doping, temperature and momentum dependence. In particular, they provide a natural explanation for the `kink puzzle': As illustrated in Fig. 3(a), the nodal kink is a result of the interband scattering on the spin-fluctuations from the upper, universal, weakly temperature-dependent branch of the spectrum ($Q_2$ vector), while the scattering between the antinodal regions ($Q_1$ vector) is determined by the middle of the spin-fluctuation spectrum where a large peak, known as a `resonance mode', appears just below $T_c$.

The determined value\footnote{For the exact formulas see \cite{DahmNP2009}, since the numeric factors are omitted here.} of $\bar{U}$ = 1.59 eV gives an estimate of $T_c$ exceeding 150 K \cite{DahmNP2009}. This demonstrates that the spin fluctuations have sufficient strength to mediate high-temperature superconductivity. The fact that the estimated value of $T_c$ is about two times higher than the measured one seems natural since the actual transition temperature can be reduced by a variety of effects. On the other hand, it is important to mention that, within the described scenario, $T_c$ should increase monotonously with underdoping, in clear contradiction to experiment. The resolution of this controversy is related to understanding of the pseudogap phenomenon, as discussed in Section \ref{sec:3}.

Having shown that the spin-fluctuation spectrum, measured by INS, describes perfectly the renormalization of the one-particle fermionic spectrum, measured by ARPES, and even can mediate high-$T_c$ superconductivity, one can follow a similar procedure described by Eq.~3 to clarify its origin. If the spin-fluctuation spectrum is formed by two-particle fermionic excitations, it should be proportional to the imaginary part of the dynamic spin susceptibility \cite{ChubukovReview}. One should note, however, that Eq.~3 gives the bare susceptibility, $\chi_0$. The dynamic one can be then derived within the random phase approximation (RPA). Indeed, the RPA spin susceptibility, modeled based on electronic band structure parameters, reproduces the low energy part of the INS spectrum, including the ($\pi,\pi$) resonance, as well as gives similar explanation for the nodal kink as produced by coupling to the upper branch of the spectrum \cite{EreminPRL2005}.

In Ref.~\cite{InosovPRB2007}, the dynamic susceptibility, for both odd and even channels, has been derived from the fermionic spectral function of BSCCO accurately mapped by ARPES. The detailed comparison with INS results supports the idea that the magnetic response below $T_c$ (or at least its major part) can be explained by the itinerant magnetism. Namely, the itinerant component of $\chi$, at least near optimal doping, has enough intensity to account for the experimentally observed magnetic resonance in both INS channels. Taking into account the out-of-plane exchange interaction, the energy difference between the odd and even resonances as well as their intensity ratio are perfectly described. Moreover, the calculated incommensurate resonance structure is similar to that observed in the INS experiment \cite{InosovPRB2007}.

\subsection{The role of phonons}
\label{sec:2.4}

One should note that similarly good description of both the nodal kink and antinodal dip has been obtained with an anisotropic coupling to two phonon modes \cite{DevereauxPRL2004}. Moreover, the strong temperature dependence of the antinodal renormalization, namely its abrupt increase below $T_c$, can be also explained as a result of the superconducting gap opening. The same can be said about other effects which we do not discuss here, the `odd scattering' \cite{Odd,OddYBCO} and the `magnetic isotope effect' \cite{TerashimaZn,Zn}. Both have been suggested as evidences for the spin-fluctuation origin of the renormalization in cuprates and later digested by the multiple phonons scenario \cite{LeePRB2007}.

In this situation, the critical (as shown in Fig.~2(d)) doping dependence of the strength of both the nodal kink \cite{LanzaraNature01,KordyukPRL2004,KordyukPRL2006} and antinodal dip \cite{BorisenkoPRL2003,KimPRL2003} remains the main problem for the phonon scenario. The idea of carrier density dependent screening of some phonon modes may explain some doping dependence of the energy scales (the positions of kink and dip) \cite{LeePRB2007,JohnstonACM2010} and only a moderate doping dependence of the renormalization strength, in an agreement with some experiments \cite{JohnstonACM2010}. In this case, one should stress the disagreement between those experiments and the experiments reviewed in the previous sections.

Having said that, we should admit that the question about the role of phonons in cuprates stays open. Here we argue that the overall fermionic spectrum, $A(\textbf{k},\omega)$, with its most remarkable features, the nodal kink and the antinodal dip, can be well described by coupling between electrons, but coupling to the lattice should also be present and is expected to be responsible for the fine structure in $A(\textbf{k},\omega)$ \cite{ZhouPRL2005,ZhaoX2010}. Moreover, revealing fine structure in ARPES spectra can essentially change our estimate for the renormalization at the very low binding energies. An example for this is recently discovered 10 meV kink \cite{Plumb}. It may appear that the interaction responsible for this kink plays an important role in superconducting pairing.

\section{Abnormal normal state}
\label{sec:3}

\begin{figure}
\center{
\resizebox{0.9\columnwidth}{!}{%
  \includegraphics{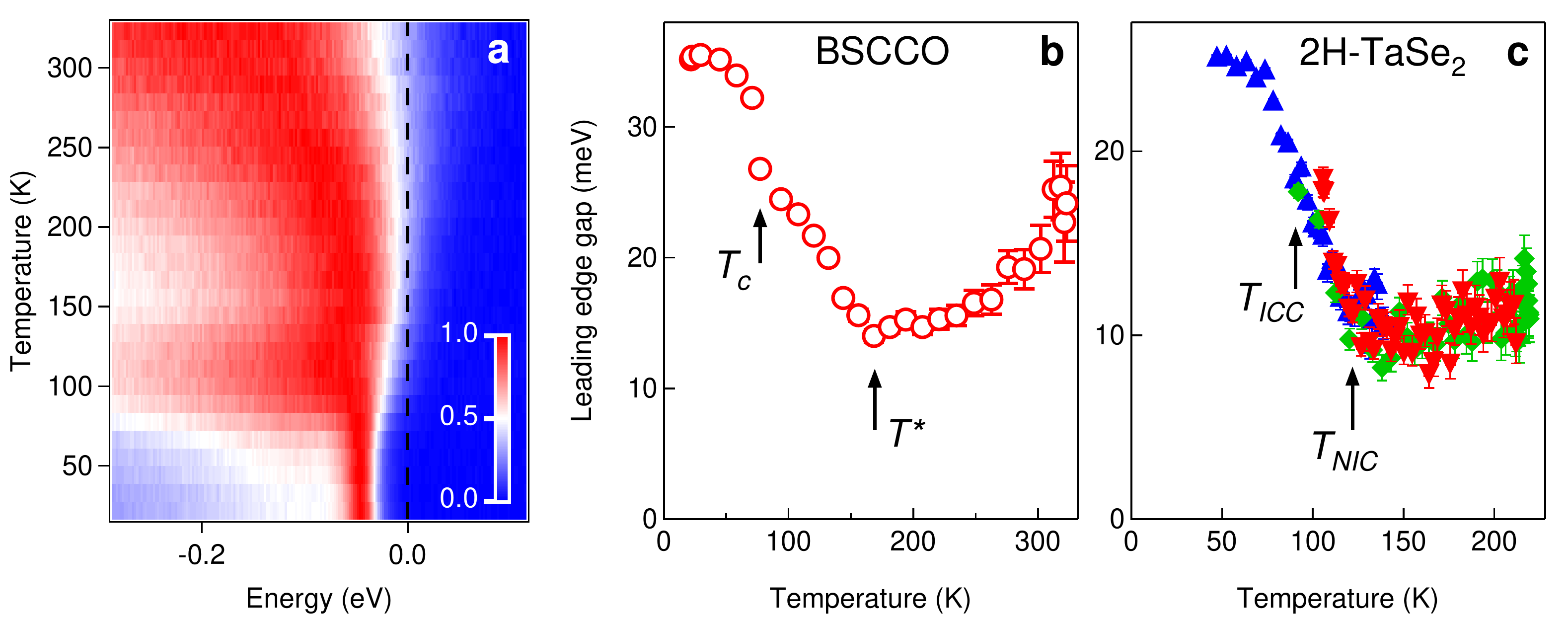} }}
\caption{Nonmonotonic pseudogap. (a) The temperature map which consists of a number of momentum integrated energy distribution curves (EDCs) measured at different temperatures at a `hot spot'. The gap is seen as a shift of the leading edge midpoint (LEM) which corresponds to white color close to the Fermi level. (b) The position of LEM as function of temperature for an underdoped Tb-BSCCO with $T_c$ = 77 K and $T^*$ = 170 K is remarkably similar to the pseudo-gap in a transition-metal dichalcogenide 2H-TaSe$_2$ (c) with the transitions to the commensurate and incommensurate CDW phases at $T_{ICC}$ = 90 K and $T_{NIC}$ = 122 K, respectively. After \cite{KordyukPG}.}
\label{fig:4}       
\end{figure}

An evident weakness of Eq.~1 is that it does not take into account the `pseudogap' (PG) \cite{TimuskPG,NormanPG}. Recent progress in ARPES measurements has led to the `two gaps' idea \cite{Tanaka2G,Kondo2G}, according to which the superconducting gap and pseudogap have different origin and compete for the phase space. At the same time, there is no general consensus on this issue \cite{NormanPG,Tanaka2G,Kondo2G,LeePG,VallaPG,HuefnerPG}. Here we argue that careful temperature- and momentum-resolved photoemission experiments suggest that the origin of the pseudogap is a spin density ordering.

In Ref.~\cite{KordyukPG} it is shown that the depletion of the spectral weight in slightly underdoped Bi(Tb)-2212 superconductor, usually associated with the pseudogap, exhibits an unexpected nonmonotonic temperature dependence: decreases linearly approaching $T^*$ at which it reveals a sharp transition but does not vanish and starts to increase gradually again at higher temperature. Fig.~4(a) shows the temperature evolution of the pseudogap as a temperature map. The gap is seen as a shift of the leading edge midpoint (LEM) of a gapped EDC. Since the leading edge of the momentum integrated EDC of the non-gapped spectrum is expected to stay at zero binding energy for any temperature \cite{VallaPG,KordyukPRB2003}, the finite shift of the LEM is a good empirical measure for a gap of unknown origin. From the presented temperature map one can see an unusual temperature evolution of the gap (in terms of the colorscale, the LEM corresponds to the white color): first it decreases with increasing temperature up to about 170 K, then it starts to increase again. The temperature dependence of the LEM is summarized in Fig. 4(b) and compared to the similar quantity measured for TaSe$_2$ (panel c), for which it is known that the pseudogap results from the incommensurate charge density wave \cite{BorisenkoTaSe2,BorisenkoNbSe2}. The observed one-to-one correspondence between the temperature dependences of the pseudogap for Bi-2212 and TaSe$_2$, which is discussed in details in \cite{KordyukPG}, suggests that density wave ordering also appears in cuprates and, reducing the electron density of states at the Fermi level, competes with superconductivity. One may assume that the spin-fluctuations, being a dominant mediator for electronic interactions in the cuprates, play also the role of the main driving force for the electronic instability resulting in the spin density wave formation. This assumption is based on the same `proximity-to-antiferromaget' argument: the increase of the pseudogap with underdoping and vanishing with overdoping.

\section{Phonons in ferropnictides}
\label{sec:4}

\begin{figure}
\center{
\resizebox{0.7\columnwidth}{!}{%
  \includegraphics{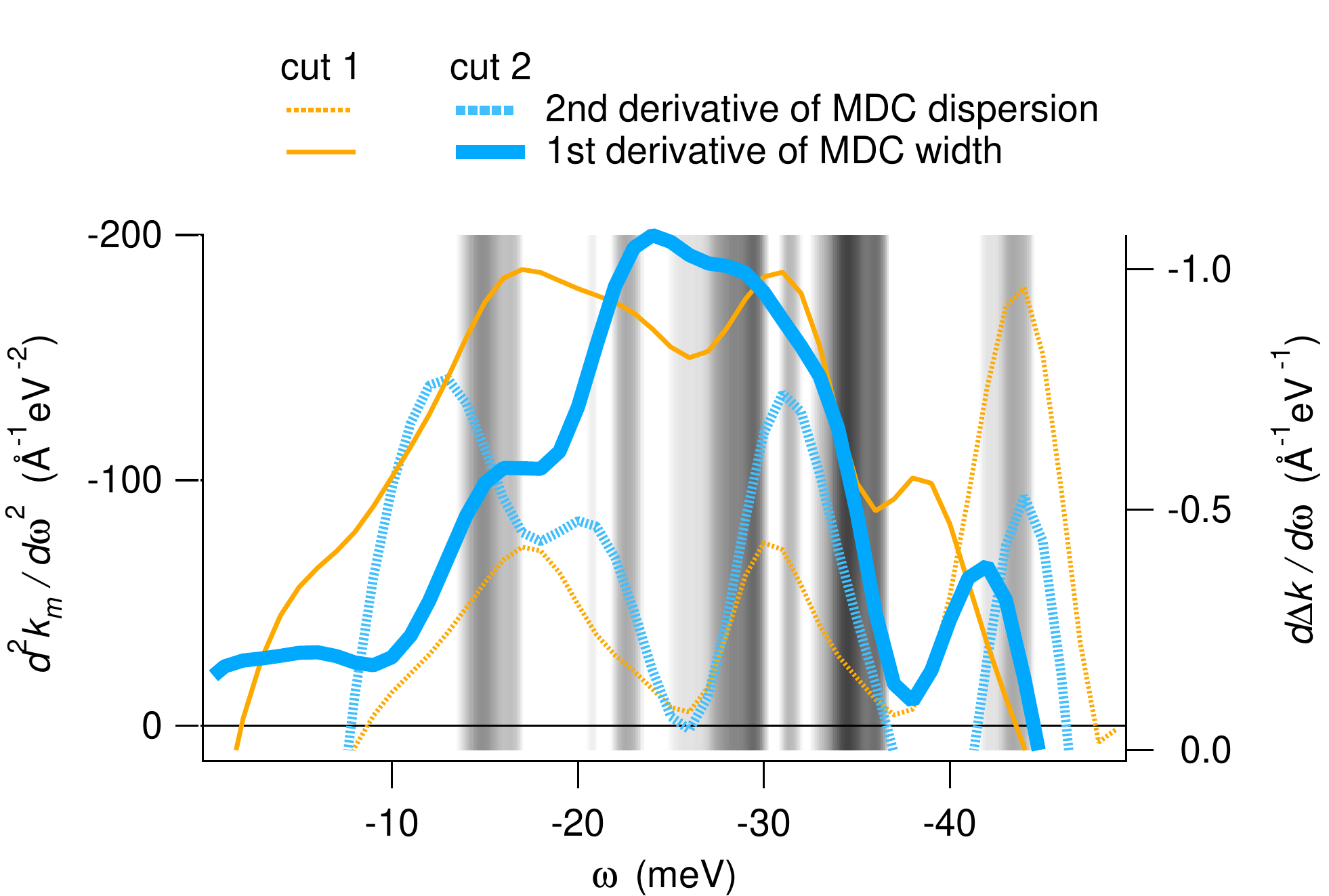} }}
\caption{Fingerprints of phonons in LiFeAs. First and second derivatives of the MDC dispersions and MDC widths, respectively, on top of the phonon density of states (gray color scale). After \cite{KordyukLFA}.}
\label{fig:5}       
\end{figure}

The iron based pnictides, a newly discovered family of high-$T_c$ superconductors, seems to provide a clear case where the phonons definitely lose vs spin-fluctuations in the nomination for the most probable pairing glue. First, the spin-fluctuations are generally expected to be strong in all the ferropnictides \cite{MazinX2009MazinNature,KorshunovPRL2009,EreminPRB2010}. Second, a number of estimates have suggested that the electron-phonon coupling there is by far not sufficient to mediate the pairing \cite{BoeriPRL2008,JishiAiCMP2010}. So, it would be interesting to apply a similar analysis of the fermionic self-energy to the pnictides.

LiFeAs seems to be a key compound for this task. While in other pnictides the self-energy analysis is complicated by essentially three-dimensional electronic band structure and magnetic ordering \cite{ZabolotnyyN2009}, LiFeAs does not show any static magnetic ordering but has rather high critical temperature ($T_c$ = 18 K) and a sizeable isotropic superconducting gap with $2\Delta/kT_c$ = 4 \cite{BorisenkoLFA,InosovLFA}. It also provides the simplest case from an experimental point of view. First, it is a stoichiometric compound that exhibits superconductivity at ambient pressure without chemical doping, thus can be easily studied by the experimental techniques which require samples without impurities. Second, although the its electronic band structure is similar to other pnictides \cite{BoeriPRL2008,JishiAiCMP2010,NekrasovJETPL2008,SinghPRB2008}, it has a perfectly two-dimensional electronic band \cite{BorisenkoLFA} well separated in momentum space from other bands. Finally, LiFeAs cleaves between the two layers of Li atoms resulting in equivalent and neutral counterparts, offering a unique opportunity to overcome the problems arising from a polar surface that can be crucial for the surface sensitive methods \cite{ZabolotnyyPRB2007}. This altogether has allowed us to derive precisely the quasiparticle self-energy from ARPES spectra and analyze its fine structure \cite{KordyukLFA}.

Fig.~5 summarizes this analysis, presenting 1st and 2nd derivatives of the MDC dispersions and MDC widths, respectively, on top of the phonon density of states calculated in \cite{JishiAiCMP2010}. The peaks on those functions are expected to coincide with peaks in the corresponding bosonic spectrum, and one can see that the correspondence is remarkable. This result, together with the estimated electron-phonon coupling strength $\lambda_{ph} = 1.38$, shows that electron-phonon coupling in pnictides is much higher than in cuprates and may be important for superconducting pairing \cite{KordyukLFA}.

\section{Conclusions}
\label{sec:5}

A careful and systematic study of fermionic spectrum of high-$T_c$ cuprates by ARPES has allowed us to derive the fermionic self-energy, analyze its structure, and identify the `fingerprints' of the spin-fluctuation spectrum. The uncovered strong dependence of the intensity of those fingerprints (of the electron-boson coupling strength) on doping and temperature unambiguously supports the magnetic origin of the key interaction. Therefore, we conclude that the spin-fluctuations play a decisive role in formation of the fermionic excitation spectrum in the normal state and are sufficient to explain the high transition temperatures to the superconducting state. The pseudogap phenomenon is consistent with this scenario and is a consequence of a Peierls-type intrinsic instability of electronic system to formation of an incommensurate spin density wave. Ironically, a similar analysis being applied to the iron pnictides reveals especially strong electron-phonon coupling that suggests important role of phonons for high-$T_c$ superconductivity in in these compounds.

\vspace{6 pt}
We acknowledge discussions with I. Eremin, A. Chubukov, I. I. Mazin, T. Dahm, D. J. Scalapino, S.-L. Drechsler, M.~L.~Kuli\'{c}, W. Hanke, B. Keimer, V. Hinkov, P. Bourges, R. Hackl, T. Valla, T. P. Devereaux, A. M. Gabovich, Yu. V. Kopaev, M. V. Sadovskii, E.~G. Maksimov, V. M. Loktev, E. A. Pashitskii, A. Semenov, J. Fink, M. Golden, M. Knupfer, A. Koitzsch, R. Schuster, and technical support from R. H\"{u}bel. The work is supported by DFG (Forschergruppe FOR538) and BMBF.

\end{document}